\documentclass[12pt]{article} 
\usepackage{times}
\usepackage{geometry}
\usepackage[utf8]{inputenc}
\usepackage{enumitem,amssymb}
\usepackage{ragged2e}
\newlist{thematic}{itemize}{8}
\setlist[thematic]{label=$\square$}
\usepackage{pifont}
\usepackage[sort, numbers]{natbib}
\usepackage{graphicx}
\usepackage{float}
\usepackage{subfig}
\usepackage{parskip}
\usepackage{multicol}

\usepackage{journals}  

\usepackage{fancyhdr}
\usepackage[yyyymmdd,hhmmss]{datetime}
\newcommand{\cmark}{\ding{51}}%
\newcommand{\done}{\rlap{$\square$}{\raisebox{2pt}{\large\hspace{1pt}\cmark}}%
\hspace{-2.5pt}}

\newcommand\arcsec{\mbox{$^{\prime\prime}$}}%

\begin{document}

\raggedright
\thispagestyle{empty}    
\huge
Astro2020 Science White Paper

\hrule
{\em\Huge Twelve Decades}: Probing the Interstellar\\ 
\vspace{-2mm}
Medium from kiloparsec to sub-AU scales
\vspace{0.5cm}

\normalsize

\noindent \textbf{Thematic Areas:} \hspace*{60pt} $\square$ Planetary Systems \hspace*{10pt} $\square$ Star and Planet Formation \hspace*{20pt}\linebreak
$\square$ Formation and Evolution of Compact Objects \hspace*{31pt} $\square$ Cosmology and Fundamental Physics \linebreak
  $\square$  Stars and Stellar Evolution \hspace*{1pt} $\done$ Resolved Stellar Populations and their Environments \hspace*{40pt} \linebreak
  $\square$    Galaxy Evolution   \hspace*{45pt} $\done$             Multi-Messenger Astronomy and Astrophysics \hspace*{65pt} \linebreak

\vspace{-2mm}
Primary thematic area: Resolved Stellar Populations and their Environments \linebreak
\vspace{-2mm}
Secondary thematic area: Multi-Messenger Astronomy and Astrophysics\linebreak
  
\vspace{-2mm}
\begin{multicols*}{2}
\raggedright
\textbf{Principal Author:}\\
\vspace{2mm}
Name: Dan R.\ {\bf Stinebring}	
 \linebreak						
Institution: Oberlin College 
 \linebreak
Email: dan.stinebring@oberlin.edu
 \linebreak
Phone:  +1 440 775 8331
\vspace{5mm}
\linebreak
\textbf{Co-authors:} (names and institutions)
\vspace{2mm}
\linebreak
   	Shami {\bf Chatterjee} \textit{(Cornell University)}\\
    Susan E.\ {\bf Clark} \textit{(Institute for Advanced Study)}\\
	James~M.~{\bf Cordes} \textit{(Cornell University)}\\
	Timothy {\bf Dolch} \textit{(Hillsdale College)}\\
	Carl {\bf Heiles} \textit{(UC Berkeley)}\\
	Alex~S.~{\bf Hill} \textit{(Univ. British Columbia}\\
	Megan {\bf Jones} \textit{(West Virginia University)}\\
	Victoria {\bf Kaspi} \textit{(McGill University)}\\
	Michael T.\ {\bf Lam} \textit{(West Virginia University)}\\
	T.\ J.\ W.\ {\bf Lazio}  \textit{(Jet Propulsion Laboratory)}\\
	Natalia {\bf Lewandowska} \textit{(West Virginia Univ.)}\\
	Dustin R.\ {\bf Madison} \textit{(West Virginia University)}\\
	Maura A.\ {\bf McLaughlin} \textit{(West Virginia Univ.)}\\
	Naomi {\bf McClure-Griffiths} \textit{(Australia National University)}\\
	Nipuni {\bf Palliyaguru} \textit{(Arecibo Observatory)}\\
	Barney J.\ {\bf Rickett} \textit{(UC San Diego)}\\
    Mayuresh P.\ {\bf Surnis} \textit{(West Virginia University)}\\
\columnbreak
\vspace*{\fill}
\fbox{\raggedright \begin{minipage}{\linewidth}
{\bf Abstract}\\ After a decade of great progress in understanding gas flow into, out of, and through the Milky Way, we are poised to merge observations with simulations to build a comprehensive picture of the multi-scale magnetized interstellar medium (ISM). These insights will also be crucial to four bold initiatives in the 2020s: gravitational waves (GWs), fast radio bursts (FRBs), cosmic B-mode, and the Event Horizon Telescope (EHT).
\end{minipage}}

\vspace{0mm}
\vspace*{\fill}
\fbox{\raggedright \begin{minipage}{\linewidth}
{\bf Related Astro2020 white papers}\\ \centerline{(by lead author(s))}\\
\\
{\bf\large NANOGrav} collaboration (J.M.\ {\bf Cordes} \& M.A.\ {\bf McLaughlin}; E.\ {\bf Fonseca}; L.Z.\ {\bf Kelley} \& M.\ {\bf Charisi}; X.\ {\bf Siemens} \& J.\ {\bf Hazboun}; S.R.\ {\bf Taylor} \& S. {\bf Burke-Spolaor}\\

Other related Astro 2020 white papers:\\
S.E.\ {\bf Clark} (Magnetic Fields and Polarization in the
Diffuse Interstellar Medium)\\
D.R.\ {\bf Lorimer} (Radio Pulsar Populations)\\
R.S.\ {\bf Lynch} (Virtues of Time and Cadence for Pulsar \& Fast Transients)\\
T.\ {\bf Bastian} (Radio Observational Constraints on Turbulent Astrophysical Plasmas)
\end{minipage}}
\vspace*{\fill}
\end{multicols*}

\pagebreak  
\pagenumbering{arabic}  
   \begin{figure*} 
    \center
    \includegraphics[width=1.0\textwidth]{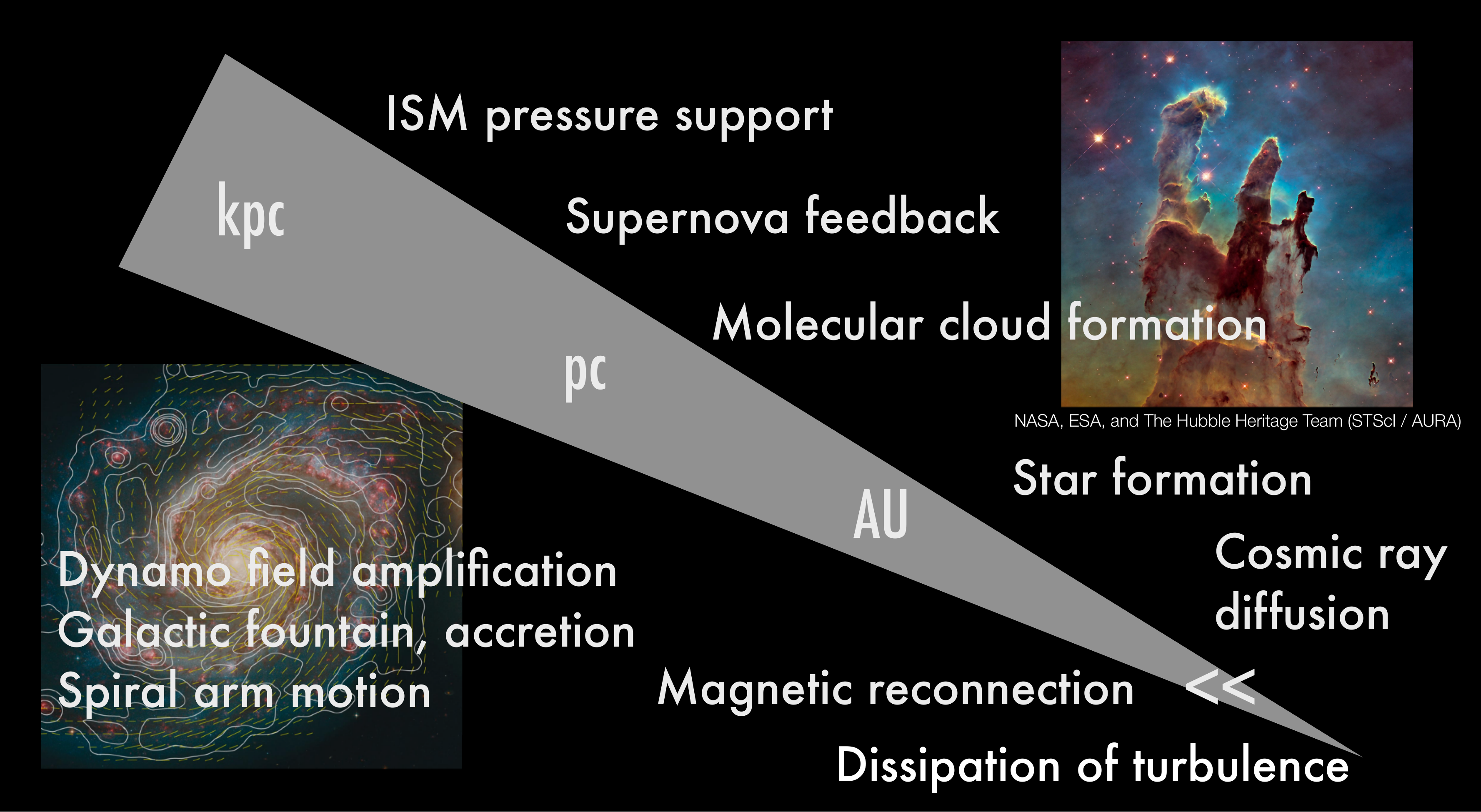}
\vspace{-4mm}
        \caption{\footnotesize The interstellar medium (ISM): from spiral arms to star formation (design: S.E. Clark; image of M51: \citep{FletcherBeck:2011})}
        \label{fig:clark}
\vspace{-3mm}
    \end{figure*}

\normalsize

\parskip 0pt

\textbf{Introduction} The interstellar gas gives birth to stars and receives their remains. Stellar ejecta both enrich it chemically and energize it by heating, ionizing, and stirring.  Thus, as illustrated in Figure 1, large-scale motions from supernovae create turbulence in the plasma, which in turn competes with gravitational contraction.  Though recent advances in indirect probing methods have revealed a cascade in energy over 12 decades in scale, many basic questions remain about the turbulence in both the neutral and ionized phases of the interstellar medium (ISM). 
We highlight this exciting and wide-ranging science here.

\vspace{4mm}
\textbf{A Decade of Challenge}%

\begin{itemize}
    \item Develop a multi-scale understanding of the ISM beyond the power spectrum.
    
    \item Incorporate magnetic structure throughout, focusing on scale-dependent dimensionality.
    
    \item Link insights probed by different tracers (including radio and optical polarization, pulsar dispersion and scintillation, H\,$\alpha$ emission, Faraday rotation, H\,\textsc{I} emission and absorption, and extinction) on radically different scales.
    
    \item Follow the energy flow from large-scale stirring to small-scale heating and cooling.
    \item Focus on a dynamical ISM and understand size-dependent timescales.
    
    \item Use the Milky Way as a guide  to extragalactic and intergalactic gas.
\end{itemize}

\vspace{3mm}
\textbf{Support Large-scale Projects}

\begin{itemize}
    \item Mitigate ISM effects and detect long-wavelength gravitational waves with a pulsar timing array (PTA).
    \item Add key propagation insights to interpret fast radio bursts (FRBs).
    \item Use neutral hydrogen mapping to better constrain the polarized light foreground, making a cosmic B-mode detection possible.
        \item Provide crucial scatter-broadening corrections for the Event Horizon Telescope (EHT).
\end{itemize}
\parskip 5pt
\pagebreak
{\LARGE\bf Probing the Structure and Energetics of the ISM}\\ 
\vspace{0.1cm}
Without a deep, consistent, multi-scale understanding of the ISM and its magnetic field we do not truly know how the Milky Way works.
If we do not fully grasp the flow of gas into, out of, and through the Milky Way, we fall short in understanding the development of other galaxies over cosmic time.
The problems continue to cascade throughout astrophysics.
While the past decade yielded great advances in our understanding of the ISM, driven by exciting new observations and a much tighter connection between data and magnetohydrodynamic (MHD) simulations, our insights are still relatively siloed and partial.

\begin{figure}[!htbp]
    \centering
    \subfloat[]{\includegraphics[width=0.40\textwidth]{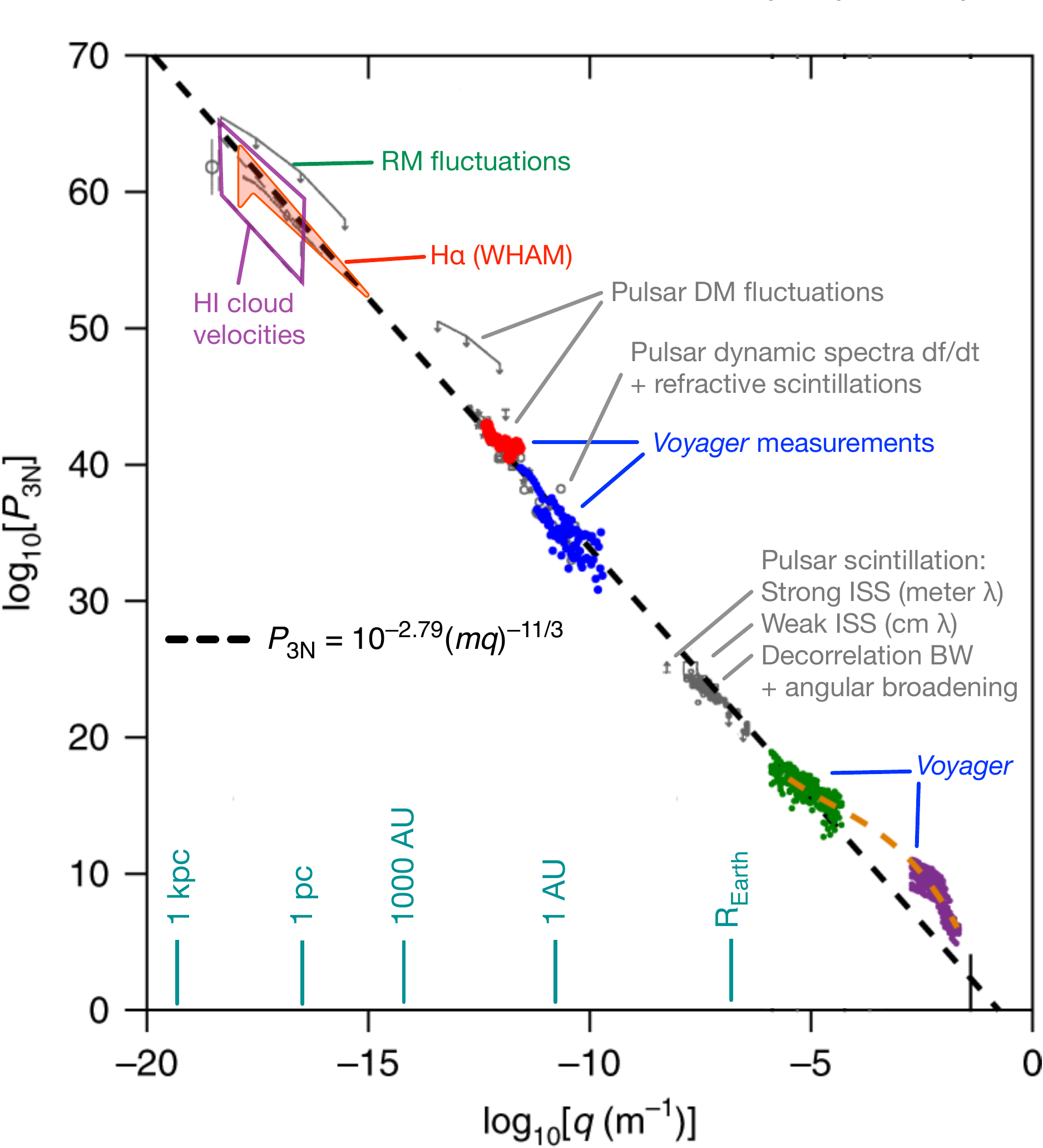}}\qquad%
    \subfloat[]{\includegraphics[width=0.34\textwidth]{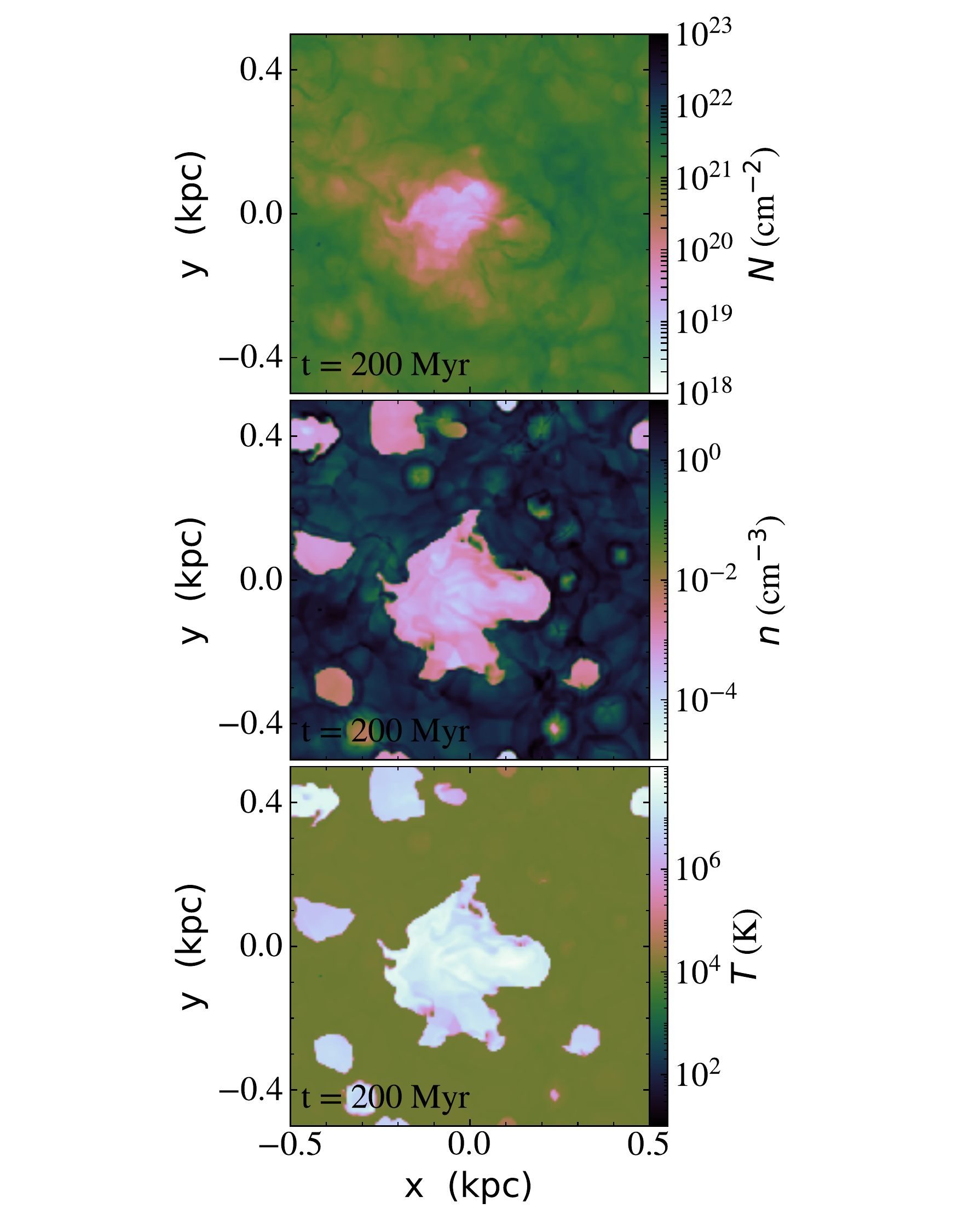}}%
        \caption{\footnotesize a) Multi-probe evidence for a Kolmogorov spectrum in plasma density, augmented recently from Voyager~1 data \citep{ars95,cl10,ll19}.  It implies a turbulent cascade over 12 decades in scale with a slight bump at the highest wavenumbers.  b) Images from a  multi-phase ISM simulation. Pinkish-white areas, some of which may be analogous to the Local Bubble, are cavities blown by supernovae \citep{hmgi18}; also, see\citep{bab09}}
        \label{fig:whythis}
    \end{figure}

As one example, in Figure~\ref{fig:whythis}a we see multi-tracer evidence for a 3-D Kolmogorov spectrum, augmented by recent {\em in situ} measurements of plasma density from the Voyager~1 spacecraft \citep{ars95,cl10,ll19}. There is enhanced power near the kinetic scales (e.\ g.\ Larmor and inertial scales). This strongly suggests a turbulent energy cascade over twelve decades in scale that is terminated by dissipation at these small scales. However, there is a disconnect between  a pervasive turbulent plasma and the presence of highly localized plasma concentrations that cause extremely anisotropic radiowave scattering or refraction (e.\ g.\  \citep{hs08}).   Pulse broadening increases strongly with distance, indicating that radiowave scattering grows very rapidly toward the inner Galaxy and is  widely distributed.  This is a problem that needs a solution. Recent Voyager~1 measurements suggest turbulence in our local region near the Sun.  
Such localized concentrations of turbulent plasma on  $\sim 1$~AU scales combine the two concepts --- appearing as spatially intermittent turbulence --- but what are the energy sources? There is no good explanation for the driver of a pervasive turbulence, nor is there an accepted mechanism for containing the thermal pressure implied in localized plasma concentrations.
MHD simulations with the needed fidelity and resolution are just arriving (e.\ g. Figure~\ref{fig:whythis}b) and will provide crucial clues for solving this puzzle.

\begin{figure}[!htbp]
    \centering
    \footnotesize
    \subfloat[]{\includegraphics[width=0.50\textwidth]{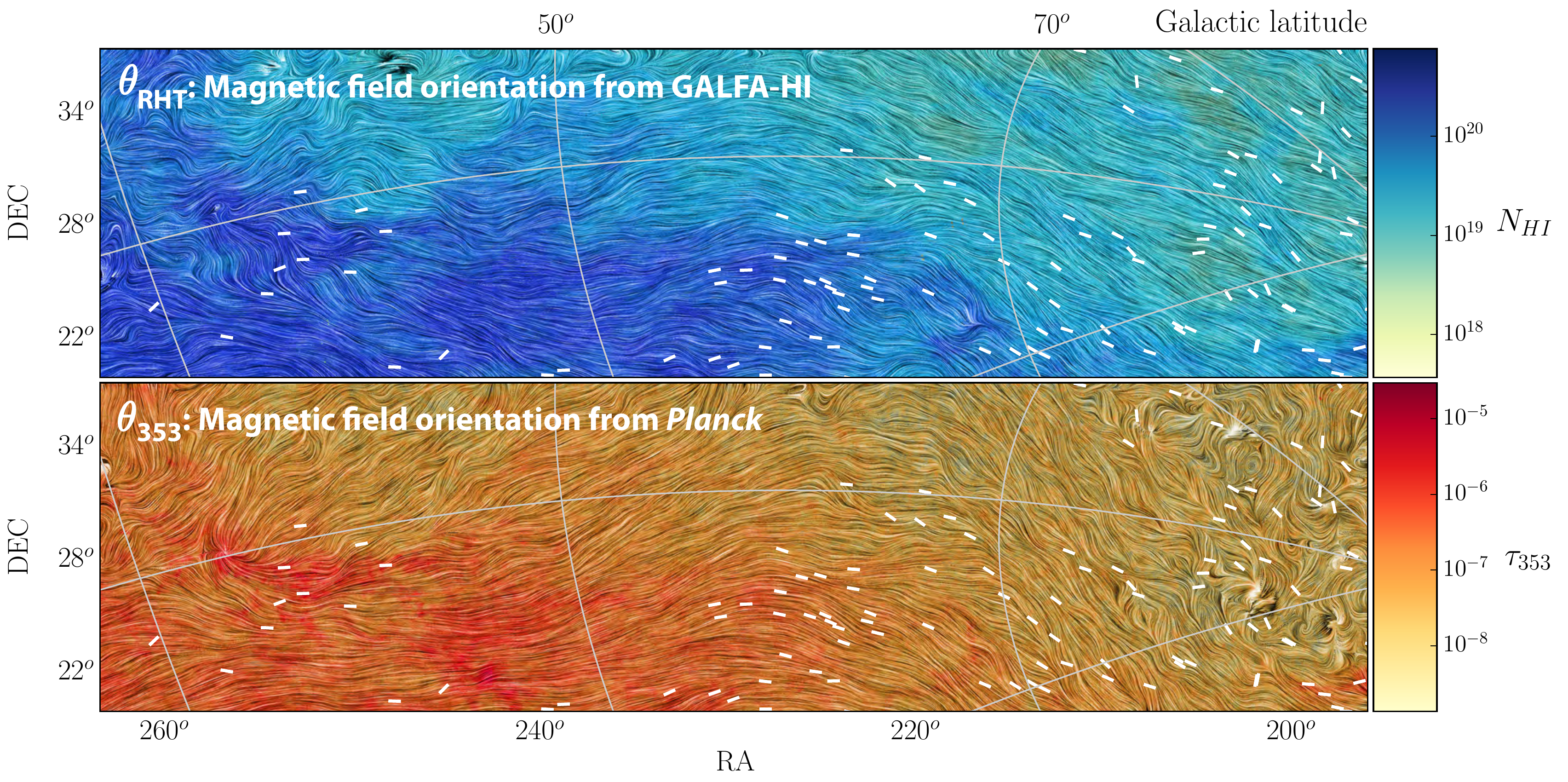}}%
    \subfloat[]{\includegraphics[width=0.50\textwidth]{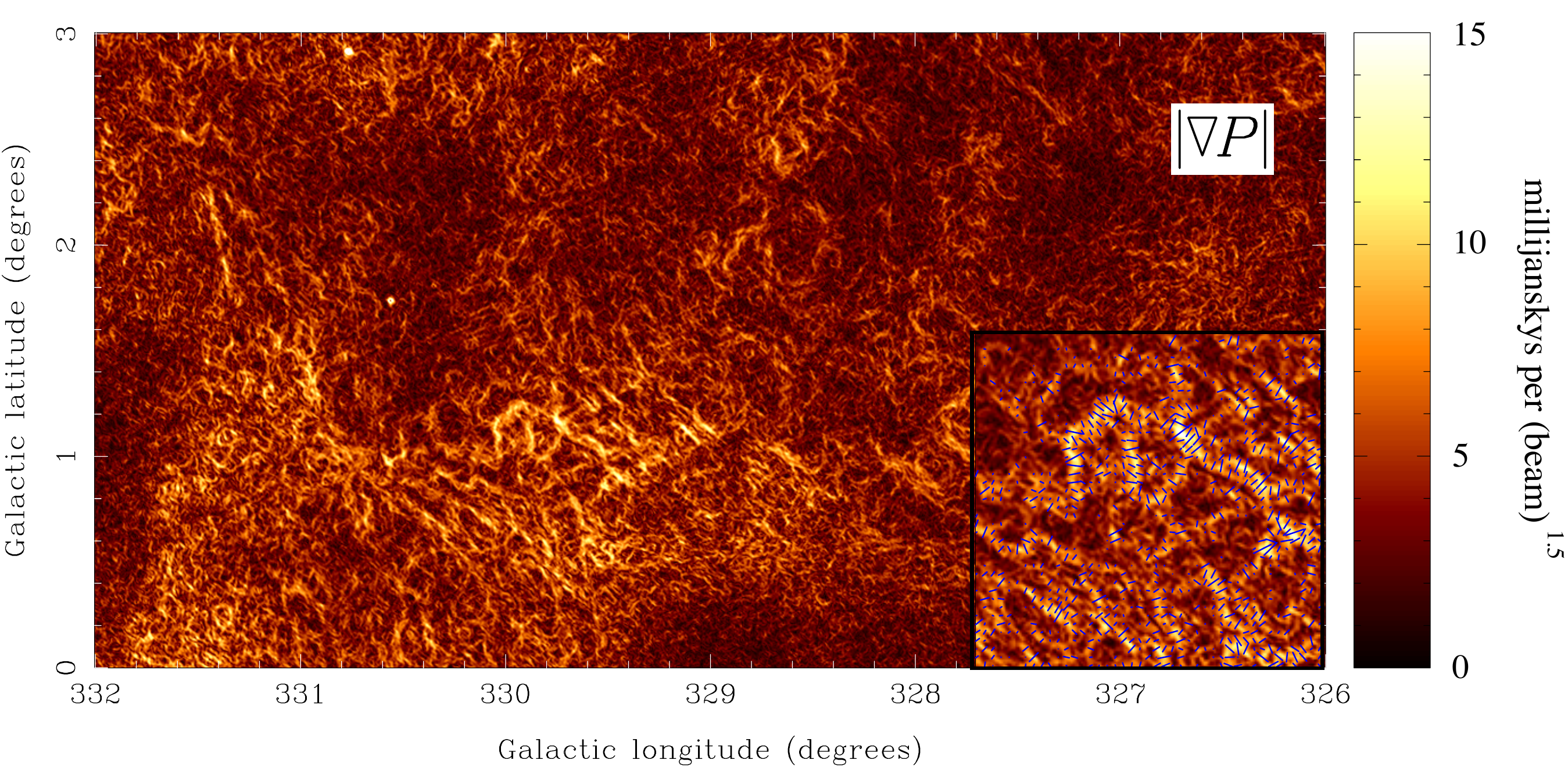}}%
        \caption{\footnotesize a) Magnetic fiber orientation from GALFA-H{\sc i} data compared with \textit{Planck} polarized dust emission \citep{cpp14,chp+15}.\quad b) Gradient image of linear polarization, $|\nabla P|$, for an 18-deg$^2$ region of the Southern Galactic Plane Survey \citep{ghb+11}.}%
        \label{fig:whynow}
    \end{figure}

\clearpage
\pagebreak
{\LARGE\bf Requirements to Achieve Scientific Goals}\\  
\vspace{0.3cm}
The last ten years have produced many surprises in our study of the gas in the Milky Way: fibers of neutral hydrogen stretching for tens of parsecs, aligned by the local magnetic field (Figure~\ref{fig:whynow}a); dense networks of magnetic field gradient (Figure~\ref{fig:whynow}b); sheets of ionized gas intercepting pulsar signals every hundred parsecs or so; compact plasma lenses of unknown origin affecting quasar and pulsar radio signals and possibly linked to FRBs. 
A comprehensive understanding of neutral and ionized gas in the Galaxy is within reach over the next decade as observations and simulations converge on a picture of turbulent, magnetized flows that yield abrupt density variations in the particular, but average, over long enough path lengths, into statistically stable turbulence cascades.

In other white papers we will highlight specific projects and facilities that will accelerate this rapid progress.
Here, we focus on the rich interlinked {\em science} that requires us to understand magnetized gas dynamics over at least twelve orders of magnitude in spatial size\footnote{For comparison, twelve orders of magnitude in  ocean size scale runs from the largest ocean waves ($\sim 30$m) to the size of an atom!}.

On the one hand there is evidence for a cascade of turbulent waves over an astonishing twelve decades in scale size. On the other hand, both in the neutral gas and in ionized hydrogen, discrete structures abound that can be explored through a variety of techniques.

What will provide science breakthroughs in the 2020s? 
The rapidly increasing sophistication of MHD modeling is starting to be up to the task by carefully incorporating the relevant physical processes and scales (e.g. \citep{bll+14,Dobbs:2015,bl16,KimOstriker:2017a,ych+18,hmgi18}). There is also promising work in developing statistical comparisons between these models and observations (e.g. \citep{BurkhartLazarian:2012,MurrayStanimirovic:2017,HerronBurkhart:2018}). However, linking the sub-AU scales on which turbulence dissipates to the Galactic scales on which it is driven remains a major numerical challenge. This will need to be coupled with continued observations that delve further into the structure of the different phases of the ISM and their interfaces.

The current generation of single-dish 21-cm surveys (GALFA-HI \citep{pbz+18}, HI4PI \citep{hbf+16}) is unlikely to be superseded for a while, which is most relevant to the diffuse, high-latitude ISM. 
However, the plane will be surveyed by interferometers, e.g.\ GASKAP. 
There will also be a major advance in the next decade in HI spin temperature measurements with SKA and its pathfinders \citep{msm+14}.
An increase in the number of HI Zeeman measurements of the field strength of similar proportions is possible (see the Astro2020 white paper by S.E.~Clark).

If molecular clouds form at the  interfaces of streams of warm gas, we need to understand how those streams form and interact \citep{msm+14,bvpk18,cgkb12,hbh+05}.

\begin{figure}[!htbp]
    \centering
    \subfloat[]{\includegraphics[width=0.60\textwidth]{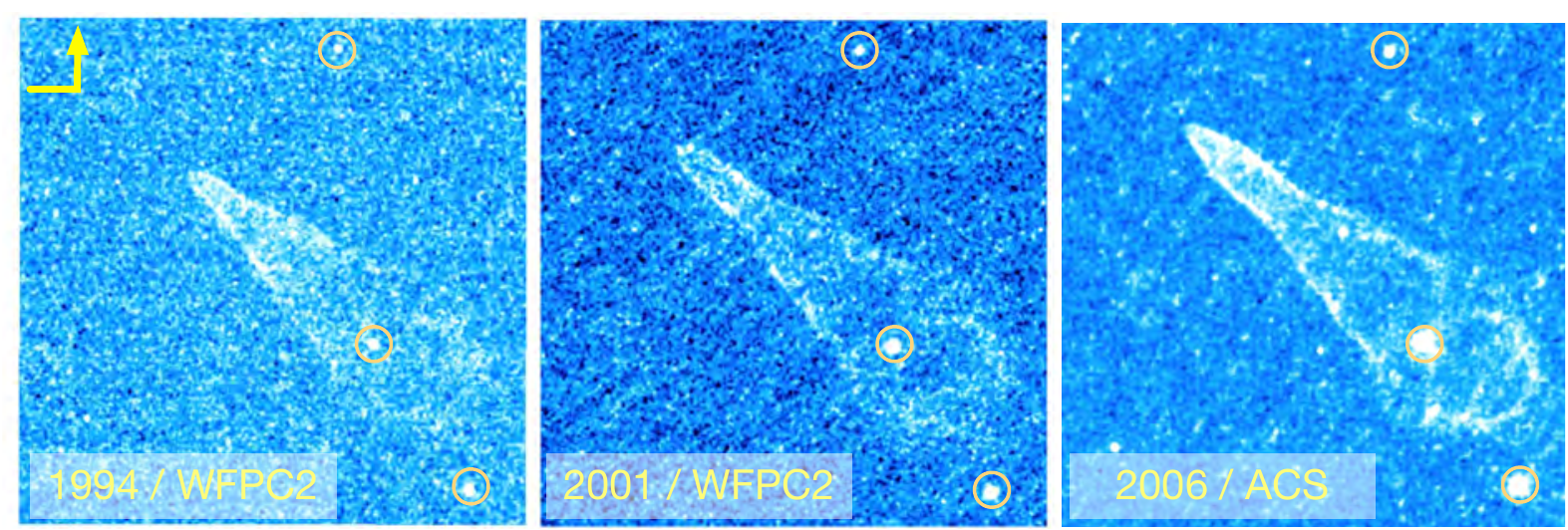}}\qquad%
    \subfloat[]{\includegraphics[width=0.30\textwidth]{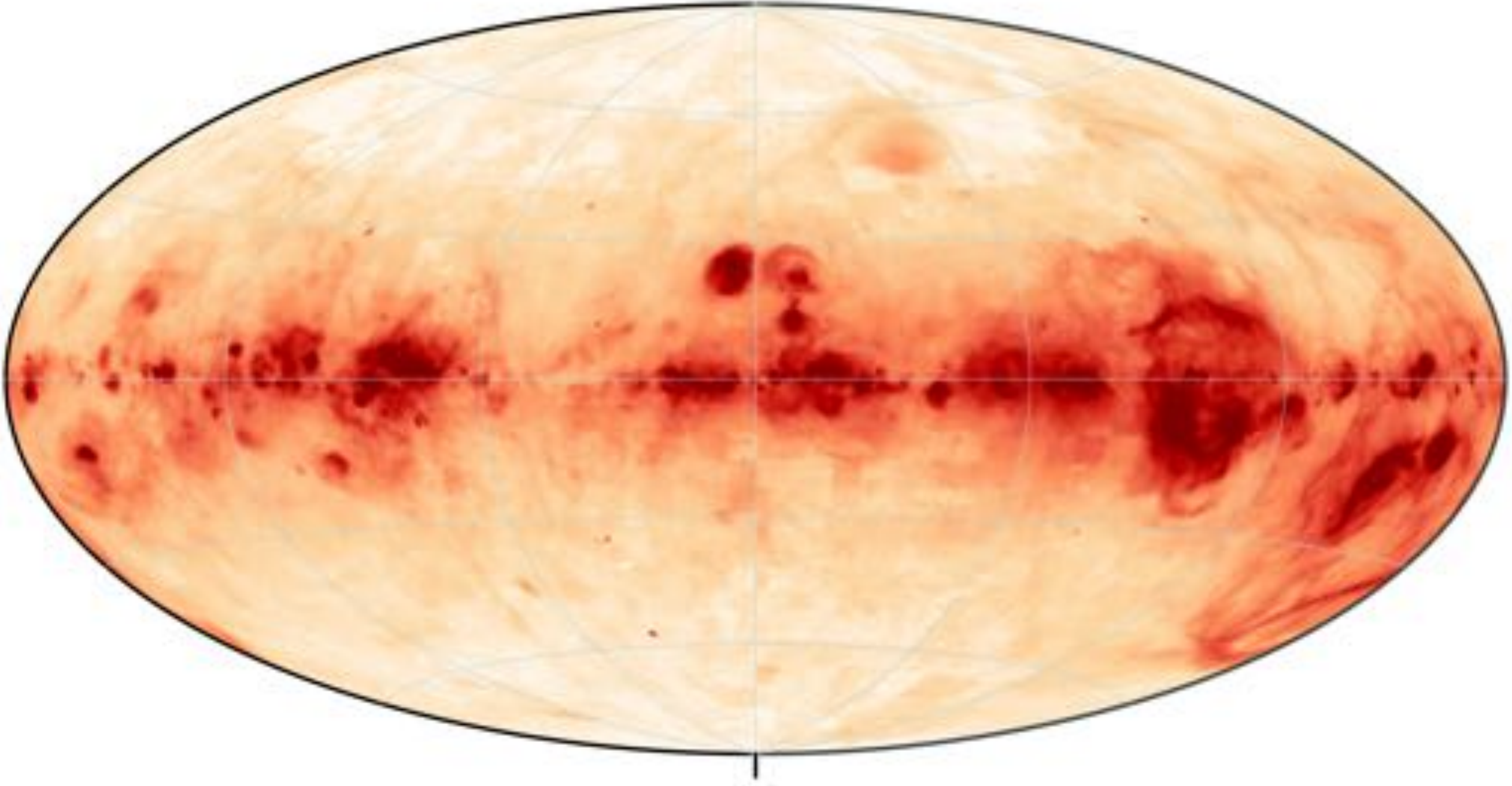}}%
        \caption{\footnotesize a) {\em HST} imaging of the time evolution of the tip of the Guitar nebula, an H$\alpha$\ bow shock nebula produced by the supersonic motion of the pulsar B2224+65, provides an {\em in situ} probe of structure in the interstellar medium. Image is 15\arcsec\ in size, and structure on 0.1\arcsec\ scales corresponds to length scales $\sim 80$~AU \citep{cc04}.\quad b) WHAM Sky Survey \citep{hrt03,hrm10}, showing all-sky H$\alpha$ emission from the WIM and H~{\sc ii} regions.}
        \label{fig:core}
    \end{figure}

As shown in Figure~\ref{fig:core} we have a rich set of diagnostics at our disposal to explore the ISM. 
 Bow-shock nebulae provide a nearly perfect {\em in situ} probe of the gas being ionized by pulsar ram pressure \citep{cc04}. Targeted H$\alpha$ observations \citep{br14} have begun to yield new ISM structures on arcsecond scales as well as constraints on warm neutral medium filling fractions. Additionally, the IPHAS \citep{iphas14} and VPHAS \citep{vphas14} surveys continue to image the H$\alpha$ distribution down to 1~arcsec along the Galactic plane, using 2--4\,m optical telescopes. 
The all-sky WHAM view of H$\alpha$ emission emphasizes the dynamic and multi-scale nature of the ionized gas \citep{hrt03,hrm10}.

\begin{figure}[!htbp] 
    \center
    \includegraphics[width=0.95\textwidth]{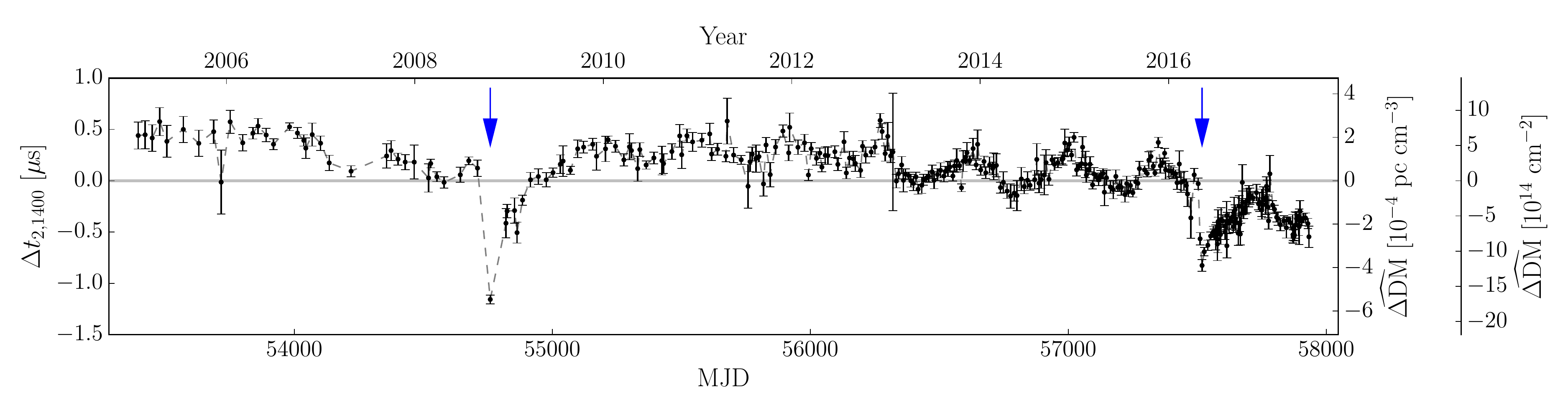}
\vspace{-4mm}
        \caption{\footnotesize Two chromatic timing events in the millisecond pulsar J1713+0747 presumably caused by  discrete ISM structures \citep{leg+18}.}
        \label{fig:J1713}
\vspace{-3mm}
    \end{figure}

High-cadence pulsar timing provides a wealth of  scientific dividends beyond the primary goals of probing neutron star physics and gravitational wave science (see the Astro2020 white paper by\\ R.\ Lynch).
As shown in Figure~\ref{fig:J1713}, the rapid space velocities of pulsars (typically 100's of km/s) scan through the ISM and uncover unexpected and unexplained structure in the parsec -- AU range. Greater observing bandwidths will also increase the range of spatial scales probed by multipath progagation \citep{CordesShannon:2016}. 
Combining scintillation studies with VLBI (``Scintillometry") can resolve scattering screens and thus improve small-scale ISM measurements \citep{popov17,mvp+17,sprb18,mvmp18,myc+18}.
      \vspace{-5mm}

\clearpage
\pagebreak
{\LARGE\bf Additional Science Dividends}\\  
 \vspace{3mm}
 In addition to  core science goals, ISM studies will benefit astrophysics in at least two other ways.
 
 First, the ISM is a serious foreground for several decade-long efforts: i) opening another gravitational wave window using a PTA, ii) unlocking the secrets of FRBs, iii) detecting a cosmic B-mode signal, and iv) imaging the Milky Way's central black hole with the EHT.
Proof-of-concept has been produced for mitigating foreground effects.
Now, we must develop robust implementation strategies, a 10-year effort in some cases.

\begin{figure}[!htbp] 
    \centering
    \subfloat[]{\includegraphics[width=0.62\textwidth]{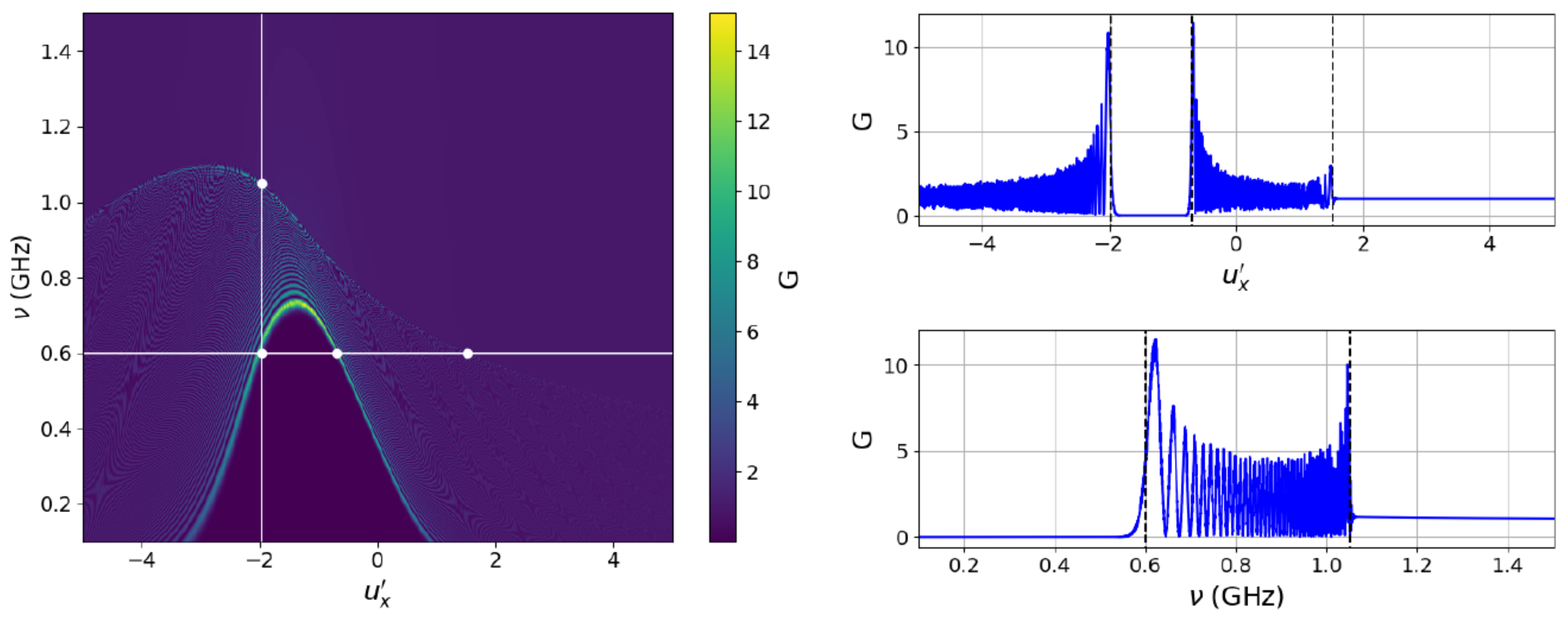}}
    \subfloat[]{\includegraphics[width=0.38\textwidth]{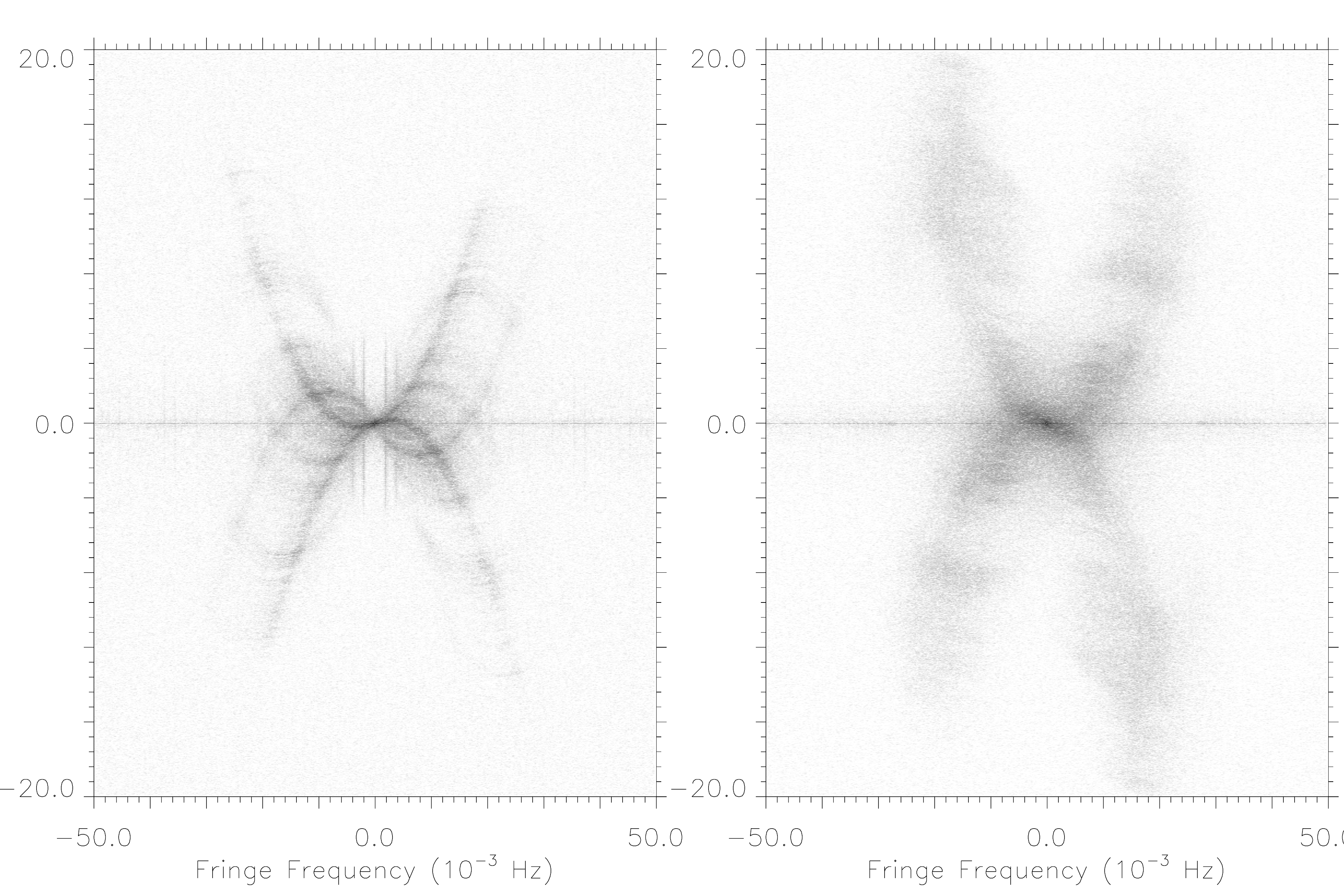}}
\vspace{-5mm}        \caption{\footnotesize a) Intensity disturbance due to lens intercept \citep{gc18}\quad b) Secondary spectra for the pulsar B1737+13 taken 6 days apart \citep{hs08}. These patterns encode the location of scattering material along the line of sight and change as the pulsar moves through the ISM at hundreds of km/s. Collectively, they indicate that intense scattering is highly localized.}

        \label{fig:additional}

    \end{figure}

Second, several high-cadence monitoring projects (PTAs, pulsar and FRB searches) produce a wealth of information about the ISM.
Synoptic telescopes like CHIME, HIRAX and DSA2000 are likely to will produce high-cadence, multifrequency data sets on many objects (pulsars and FRBs), yielding important ISM information.
In the case of PTAs, high-precision ionized column densities and scatter-delay measurements are made weekly or biweekly.
Figures~\ref{fig:J1713} and  \ref{fig:additional} show some of the unexpected events that have occurred or may appear in this rich data set, which will grow to about 100 sight lines through the Galaxy in the 2020s, sampled as often as daily.
The ability to detect and rapidly follow up on ISM events will complement other synoptic monitoring efforts such as the LSST.
In Figure~6b we show two scintillation arc observations taken 6~days apart. 
This technique has led to the realization that along many sight lines the scattering is highly localized, persistent in basic ``screen" location, but highly variable (``intermittent") as the sight line scans transversely through the ISM \citep{smc+01,wmsz04,hsa+05,crsc06,hs07,wksv08,rscg11,css16,sro19} 

\vspace{5mm}
\noindent{\Large\bf Summary}

Surprises from the last ten years emphasize that the ISM is a multi-scale medium threaded with tangled and ordered magnetic fields that dominate its dynamics in many cases.
Continued support of this science in the next ten years will {\bf yield crucial payoffs for US and international astrophysics:} from the interaction with star formation, not discussed here; through the gas dynamics of the Milky Way including numerous episodes of large-scale infall and expulsion; and by the feedback mechanisms associated with supernovae and powerful stellar winds.
Knowing the full energy budget of the Milky Way leads to clear insights about external galaxies and galaxy assembly through cosmic time.
In addition, the science sketched here provides crucial support to four breakthrough-oriented initiatives of the 2020s.

  \vspace{-5mm}

\clearpage
\pagebreak


\begin{thebibliography}{10}

\bibitem{FletcherBeck:2011}
Andrew Fletcher, R~Beck, A~Shukurov, E~M Berkhuijsen, and C~Horellou.
\newblock {Magnetic fields and spiral arms in the galaxy M51}.
\newblock {\em MNRAS}, 412(4):2396--2416, April 2011.

\bibitem{ars95}
J.~W. Armstrong, B.~J. Rickett, and S.~R. Spangler.
\newblock Electron density power spectrum in the local interstellar medium.
\newblock {\em \apj}, 443:209--221, 1995.

\bibitem{cl10}
A.~{Chepurnov} and A.~{Lazarian}.
\newblock {Extending the Big Power Law in the Sky with Turbulence Spectra from
  Wisconsin H{\ensuremath{\alpha}} Mapper Data}.
\newblock {\em \apj}, 710:853--858, Feb 2010.

\bibitem{ll19}
K.~H. Lee and L.~C. Lee.
\newblock Interstellar turbulence spectrum from {\em in situ} observations of
  voyager 1.
\newblock {\em \natast}, 3:154--159, 2019.

\bibitem{hs08}
D.~A. Hemberger and D.~R. Stinebring.
\newblock {Time Variability of Interstellar Scattering and Improvements to
  Pulsar Timing}.
\newblock {\em \apj}, 674:L37--L40, 2008.

\bibitem{hmgi18}
Alex~S. {Hill}, Mordecai-Mark {Mac Low}, Andrea {Gatto}, and Juan~C.
  {Ib{\'a}{\~n}ez-Mej{\'\i}a}.
\newblock {Effect of the Heating Rate on the Stability of the Three-phase
  Interstellar Medium}.
\newblock {\em \apj}, 862:55, Jul 2018.

\bibitem{bab09}
Dieter {Breitschwerdt}, Miguel~A. {de Avillez}, and Verena {Baumgartner}.
\newblock {Modeling the Local Warm/Hot Bubble}.
\newblock In Randall~K. {Smith}, Steven~L. {Snowden}, and K.~D. {Kuntz},
  editors, {\em American Institute of Physics Conference Series}, volume 1156
  of {\em American Institute of Physics Conference Series}, pages 271--279, Aug
  2009.

\bibitem{cpp14}
S.~E. {Clark}, J.~E.~G. {Peek}, and M.~E. {Putman}.
\newblock {Magnetically Aligned H I Fibers and the Rolling Hough Transform}.
\newblock {\em \apj}, 789:82, Jul 2014.

\bibitem{chp+15}
S.~E. {Clark}, J.~Colin {Hill}, J.~E.~G. {Peek}, M.~E. {Putman}, and B.~L.
  {Babler}.
\newblock {Neutral Hydrogen Structures Trace Dust Polarization Angle:
  Implications for Cosmic Microwave Background Foregrounds}.
\newblock {\em \prl}, 115:241302, Dec 2015.

\bibitem{ghb+11}
B.~M. {Gaensler}, M.~{Haverkorn}, B.~{Burkhart}, K.~J. {Newton-McGee}, R.~D.
  {Ekers}, A.~{Lazarian}, N.~M. {McClure-Griffiths}, T.~{Robishaw}, J.~M.
  {Dickey}, and A.~J. {Green}.
\newblock {Low-Mach-number turbulence in interstellar gas revealed by radio
  polarization gradients}.
\newblock {\em \nat}, 478:214--217, Oct 2011.

\bibitem{bll+14}
Blakesley {Burkhart}, A.~{Lazarian}, I.~C. {Le{\~a}o}, J.~R. {de Medeiros}, and
  A.~{Esquivel}.
\newblock {Measuring the Alfv{\'e}nic Nature of the Interstellar Medium:
  Velocity Anisotropy Revisited}.
\newblock {\em \apj}, 790:130, Aug 2014.

\bibitem{Dobbs:2015}
Clare~L. {Dobbs}.
\newblock {The interstellar medium and star formation on kpc size scales}.
\newblock {\em \mnras}, 447:3390--3401, Mar 2015.

\bibitem{bl16}
Blakesley {Burkhart} and A.~{Lazarian}.
\newblock {The Phase Coherence of Interstellar Density Fluctuations}.
\newblock {\em \apj}, 827:26, Aug 2016.

\bibitem{KimOstriker:2017a}
Chang-Goo Kim and Eve~C Ostriker.
\newblock {Three-phase Interstellar Medium in Galaxies Resolving Evolution with
  Star Formation and Supernova Feedback (TIGRESS): Algorithms, Fiducial Model,
  and Convergence}.
\newblock {\em ApJ}, 846(2):133, September 2017.

\bibitem{ych+18}
Ka~Ho {Yuen}, Junda {Chen}, Yue {Hu}, Ka~Wai {Ho}, A.~{Lazarian}, Victor
  {Lazarian}, Bo~{Yang}, Blakesley {Burkhart}, Caio {Correia}, Jungyeon {Cho},
  Bruno {Canto}, and J.~R. {De Medeiros}.
\newblock {Statistical Tracing of Magnetic Fields: Comparing and Improving the
  Techniques}.
\newblock {\em \apj}, 865:54, Sep 2018.

\bibitem{BurkhartLazarian:2012}
Blakesley Burkhart, Alex Lazarian, and Bryan~M Gaensler.
\newblock {Properties of Interstellar Turbulence from Gradients of Linear
  Polarization Maps}.
\newblock {\em ApJ}, 749(2):145, April 2012.

\bibitem{MurrayStanimirovic:2017}
Claire~E. {Murray}, Sne{\v{z}}ana {Stanimirovi{\'c}}, Chang-Goo {Kim}, Eve~C.
  {Ostriker}, Robert~R. {Lindner}, Carl {Heiles}, John~M. {Dickey}, and Brian
  {Babler}.
\newblock {Recovering Interstellar Gas Properties with Hi Spectral Lines: A
  Comparison between Synthetic Spectra and 21-SPONGE}.
\newblock {\em \apj}, 837:55, Mar 2017.

\bibitem{HerronBurkhart:2018}
C~A Herron, Blakesley Burkhart, Bryan~M Gaensler, G~F Lewis, Naomi~M
  McClure-Griffiths, G~Bernardi, E~Carretti, M~Haverkorn, M~Kesteven, S~Poppi,
  and Lister Staveley-Smith.
\newblock {Advanced Diagnostics for the Study of Linearly Polarized Emission.
  II. Application to Diffuse Interstellar Radio Synchrotron Emission}.
\newblock {\em ApJ}, 855(1):29, March 2018.

\bibitem{pbz+18}
J.~E.~G. {Peek}, Brian~L. {Babler}, Yong {Zheng}, S.~E. {Clark}, Kevin~A.
  {Douglas}, Eric~J. {Korpela}, M.~E. {Putman}, Sne{\v{z}}ana
  {Stanimirovi{\'c}}, Steven~J. {Gibson}, and Carl {Heiles}.
\newblock {The GALFA-H I Survey Data Release 2}.
\newblock {\em The Astrophysical Journal Supplement Series}, 234:2, Jan 2018.

\bibitem{hbf+16}
{HI4PI Collaboration}, N.~{Ben Bekhti}, L.~{Fl{\"o}er}, R.~{Keller}, J.~{Kerp},
  D.~{Lenz}, B.~{Winkel}, J.~{Bailin}, M.~R. {Calabretta}, L.~{Dedes}, H.~A.
  {Ford}, B.~K. {Gibson}, U.~{Haud}, S.~{Janowiecki}, P.~M.~W. {Kalberla},
  F.~J. {Lockman}, N.~M. {McClure-Griffiths}, T.~{Murphy}, H.~{Nakanishi},
  D.~J. {Pisano}, and L.~{Staveley-Smith}.
\newblock {HI4PI: A full-sky H I survey based on EBHIS and GASS}.
\newblock {\em \aap}, 594:A116, Oct 2016.

\bibitem{msm+14}
N.~M. {McClure-Griffiths}, S.~{Stanimirovic}, C.~{Murray}, D.~{Li}, J.~M.
  {Dickey}, E.~{Vazquez-Semadeni}, J.~E.~G. {Peek}, M.~{Putman}, S.~E. {Clark},
  M.~A. {Miville-Deschenes}, J.~{Bland-Hawthorn}, and L.~{Staveley-Smith}.
\newblock {Galactic and Magellanic Evolution with the SKA}.
\newblock In {\em Advancing Astrophysics with the Square Kilometre Array
  (AASKA14)}, page 130, Apr 2015.

\bibitem{bvpk18}
Javier {Ballesteros-Paredes}, Enrique {V{\'a}zquez-Semadeni}, Aina {Palau}, and
  Ralf~S. {Klessen}.
\newblock {Gravity or turbulence? - IV. Collapsing cores in out-of-virial
  disguise}.
\newblock {\em \mnras}, 479:2112--2125, Sep 2018.

\bibitem{cgkb12}
Paul~C. {Clark}, Simon C.~O. {Glover}, Ralf~S. {Klessen}, and Ian~A. {Bonnell}.
\newblock {How long does it take to form a molecular cloud?}
\newblock {\em \mnras}, 424:2599--2613, Aug 2012.

\bibitem{hbh+05}
Fabian {Heitsch}, Andreas {Burkert}, Lee~W. {Hartmann}, Adrianne~D. {Slyz}, and
  Julien E.~G. {Devriendt}.
\newblock {Formation of Structure in Molecular Clouds: A Case Study}.
\newblock {\em \apj}, 633:L113--L116, Nov 2005.

\bibitem{cc04}
S.~{Chatterjee} and J.~M. {Cordes}.
\newblock {Smashing the Guitar: An Evolving Neutron Star Bow Shock}.
\newblock {\em \apj}, 600:L51--L54, Jan 2004.

\bibitem{hrt03}
L.~M. {Haffner}, R.~J. {Reynolds}, S.~L. {Tufte}, G.~J. {Madsen}, K.~P.
  {Jaehnig}, and J.~W. {Percival}.
\newblock {The Wisconsin H{$\alpha$} Mapper Northern Sky Survey}.
\newblock {\em \apjs}, 149:405--422, December 2003.

\bibitem{hrm10}
L.~M. {Haffner}, R.~J. {Reynolds}, G.~J. {Madsen}, A.~S. {Hill}, K.~A.
  {Barger}, K.~P. {Jaehnig}, E.~J. {Mierkiewicz}, J.~W. {Percival}, and
  N.~{Chopra}.
\newblock {Early Results from the Wisconsin H-Alpha Mapper Southern Sky
  Survey}.
\newblock In R.~{Kothes}, T.~L. {Landecker}, and A.~G. {Willis}, editors, {\em
  The Dynamic Interstellar Medium: A Celebration of the Canadian Galactic Plane
  Survey}, volume 438 of {\em Astronomical Society of the Pacific Conference
  Series}, page 388, December 2010.

\bibitem{br14}
S.~{Brownsberger} and R.~W. {Romani}.
\newblock {A Survey for H{$\alpha$} Pulsar Bow Shocks}.
\newblock {\em \apj}, 784:154, April 2014.

\bibitem{iphas14}
G.~{Barentsen}, H.~J. {Farnhill}, J.~E. {Drew}, E.~A. {Gonz{\'a}lez-Solares},
  R.~{Greimel}, M.~J. {Irwin}, B.~{Miszalski}, C.~{Ruhland}, P.~{Groot},
  A.~{Mampaso}, S.~E. {Sale}, A.~A. {Henden}, A.~{Aungwerojwit}, M.~J.
  {Barlow}, P.~J. {Carter}, R.~L.~M. {Corradi}, J.~J. {Drake},
  J.~{Eisl{\"o}ffel}, J.~{Fabregat}, B.~T. {G{\"a}nsicke}, N.~P. {Gentile
  Fusillo}, S.~{Greiss}, A.~S. {Hales}, S.~{Hodgkin}, L.~{Huckvale},
  J.~{Irwin}, R.~{King}, C.~{Knigge}, T.~{Kupfer}, E.~{Lagadec}, D.~J.
  {Lennon}, J.~R. {Lewis}, M.~{Mohr-Smith}, R.~A.~H. {Morris}, T.~{Naylor},
  Q.~A. {Parker}, S.~{Phillipps}, S.~{Pyrzas}, R.~{Raddi}, G.~H.~A. {Roelofs},
  P.~{Rodr{\'{\i}}guez-Gil}, L.~{Sabin}, S.~{Scaringi}, D.~{Steeghs},
  J.~{Suso}, R.~{Tata}, Y.~C. {Unruh}, J.~{van Roestel}, K.~{Viironen}, J.~S.
  {Vink}, N.~A. {Walton}, N.~J. {Wright}, and A.~A. {Zijlstra}.
\newblock {The second data release of the INT Photometric H{$\alpha$} Survey of
  the Northern Galactic Plane (IPHAS DR2)}.
\newblock {\em \mnras}, 444:3230--3257, November 2014.

\bibitem{vphas14}
J.~E. {Drew}, E.~{Gonzalez-Solares}, R.~{Greimel}, M.~J. {Irwin},
  A.~{K{\"u}pc{\"u} Yoldas}, J.~{Lewis}, G.~{Barentsen}, J.~{Eisl{\"o}ffel},
  H.~J. {Farnhill}, W.~E. {Martin}, J.~R. {Walsh}, N.~A. {Walton},
  M.~{Mohr-Smith}, R.~{Raddi}, S.~E. {Sale}, N.~J. {Wright}, P.~{Groot}, M.~J.
  {Barlow}, R.~L.~M. {Corradi}, J.~J. {Drake}, J.~{Fabregat}, D.~J. {Frew},
  B.~T. {G{\"a}nsicke}, C.~{Knigge}, A.~{Mampaso}, R.~A.~H. {Morris},
  T.~{Naylor}, Q.~A. {Parker}, S.~{Phillipps}, C.~{Ruhland}, D.~{Steeghs},
  Y.~C. {Unruh}, J.~S. {Vink}, R.~{Wesson}, and A.~A. {Zijlstra}.
\newblock {The VST Photometric H{$\alpha$} Survey of the Southern Galactic
  Plane and Bulge (VPHAS+)}.
\newblock {\em \mnras}, 440:2036--2058, May 2014.

\bibitem{CordesShannon:2016}
J~M Cordes, R~M Shannon, and D~R Stinebring.
\newblock {Frequency-dependent Dispersion Measures and Implications for Pulsar
  Timing}.
\newblock {\em ApJ}, 817(1):16, January 2016.

\bibitem{popov17}
Mikhail~V. {Popov}, Norbert {Bartel}, Carl~R. {Gwinn}, Michael~D. {Johnson},
  Andrey {Andrianov}, Evgeny {Fadeev}, Bhal~Chandra {Joshi}, Nikolay
  {Kardashev}, Ramesh {Karuppusamy}, Yuri~Y. {Kovalev}, Michael {Kramer},
  Alexey {Rudnitskiy}, Vladimir {Shishov}, Tatiana {Smirnova}, Vladimir~A.
  {Soglasnov}, and J.~Anton {Zensus}.
\newblock {PSR B0329+54: substructure in the scatter-broadened image discovered
  with RadioAstron on baselines up to 330 000 km}.
\newblock {\em \mnras}, 465:978--985, Feb 2017.

\bibitem{mvp+17}
Robert {Main}, Marten {van Kerkwijk}, Ue-Li {Pen}, Nikhil {Mahajan}, and Keith
  {Vanderlinde}.
\newblock {Descattering of Giant Pulses in PSR B1957+20}.
\newblock {\em \apj}, 840:L15, May 2017.

\bibitem{sprb18}
Dana {Simard}, Ue-Li {Pen}, Visweshwar {Ram Marthi}, and Walter {Brisken}.
\newblock {Disentangling interstellar plasma screens with pulsar VLBI:
  Combining auto- and cross-correlations}.
\newblock {\em arXiv e-prints}, page arXiv:1810.07231, Oct 2018.

\bibitem{mvmp18}
Nikhil {Mahajan}, Marten~H. {van Kerkwijk}, Robert {Main}, and Ue-Li {Pen}.
\newblock {Mode Changing and Giant Pulses in the Millisecond Pulsar PSR
  B1957+20}.
\newblock {\em \apj}, 867:L2, Nov 2018.

\bibitem{myc+18}
Robert {Main}, I.~Sheng {Yang}, Victor {Chan}, Dongzi {Li}, Fang~Xi {Lin},
  Nikhil {Mahajan}, Ue-Li {Pen}, Keith {Vanderlinde}, and Marten~H. {van
  Kerkwijk}.
\newblock {Pulsar emission amplified and resolved by plasma lensing in an
  eclipsing binary}.
\newblock {\em \nat}, 557:522--525, May 2018.

\bibitem{leg+18}
M.~T. {Lam}, J.~A. {Ellis}, G.~{Grillo}, M.~L. {Jones}, J.~S. {Hazboun}, P.~R.
  {Brook}, J.~E. {Turner}, S.~{Chatterjee}, J.~M. {Cordes}, T.~J.~W. {Lazio},
  M.~E. {DeCesar}, Z.~{Arzoumanian}, H.~{Blumer}, H.~T. {Cromartie}, P.~B.
  {Demorest}, T.~{Dolch}, R.~D. {Ferdman}, E.~C. {Ferrara}, E.~{Fonseca},
  N.~{Garver-Daniels}, P.~A. {Gentile}, V.~{Gupta}, D.~R. {Lorimer}, R.~S.
  {Lynch}, D.~R. {Madison}, M.~A. {McLaughlin}, C.~{Ng}, D.~J. {Nice}, T.~T.
  {Pennucci}, S.~M. {Ransom}, R.~{Spiewak}, I.~H. {Stairs}, D.~R. {Stinebring},
  K.~{Stovall}, J.~K. {Swiggum}, S.~J. {Vigeland}, and W.~W. {Zhu}.
\newblock {A Second Chromatic Timing Event of Interstellar Origin toward PSR
  J1713+0747}.
\newblock {\em \apj}, 861:132, Jul 2018.

\bibitem{smc+01}
D.~R. {Stinebring}, M.~A. {McLaughlin}, J.~M. {Cordes}, K.~M. {Becker},
  J.~E.~E. {Goodman}, M.~A. {Kramer}, J.~L. {Sheckard}, and C.~T. {Smith}.
\newblock {Faint Scattering Around Pulsars: Probing the Interstellar Medium on
  Solar System Size Scales}.
\newblock 549:L97--L100, 2001.

\bibitem{wmsz04}
M.~A. {Walker}, D.~B. {Melrose}, D.~R. {Stinebring}, and C.~M. {Zhang}.
\newblock {Interpretation of parabolic arcs in pulsar secondary spectra}.
\newblock {\em MNRAS}, 354:43--54, October 2004.

\bibitem{hsa+05}
A.~S. {Hill}, D.~R. {Stinebring}, C.~T. {Asplund}, D.~E. {Berwick}, W.~B.
  {Everett}, and N.~R. {Hinkel}.
\newblock {Deflection of Pulsar Signal Reveals Compact Structures in the
  Galaxy}.
\newblock {\em ApJL}, 619:L171--L174, February 2005.

\bibitem{crsc06}
J.~M. {Cordes}, B.~J. {Rickett}, D.~R. {Stinebring}, and W.~A. {Coles}.
\newblock {Theory of Parabolic Arcs in Interstellar Scintillation Spectra}.
\newblock 637:346--365, January 2006.

\bibitem{hs07}
C.~{Heiles} and D.~{Stinebring}.
\newblock {Observational Review of this SINS Meeting}.
\newblock In M.~{Haverkorn} and W.~M. {Goss}, editors, {\em SINS - Small
  Ionized and Neutral Structures in the Diffuse Interstellar Medium}, volume
  365 of {\em Astronomical Society of the Pacific Conference Series}, page 331,
  July 2007.

\bibitem{wksv08}
M.~A. {Walker}, L.~V.~E. {Koopmans}, D.~R. {Stinebring}, and W.~{van Straten}.
\newblock {Interstellar holography}.
\newblock {\em MNRAS}, 388:1214--1222, August 2008.

\bibitem{rscg11}
B.~{Rickett}, D.~{Stinebring}, B.~{Coles}, and J.~{Gao}.
\newblock {Pulsar Scintillation Arcs reveal filaments in the Interstellar
  Plasma}.
\newblock In M.~{Burgay}, N.~{D'Amico}, P.~{Esposito}, A.~{Pellizzoni}, and
  A.~{Possenti}, editors, {\em American Institute of Physics Conference
  Series}, volume 1357 of {\em American Institute of Physics Conference
  Series}, pages 97--100, August 2011.

\bibitem{css16}
J.~M. {Cordes}, R.~M. {Shannon}, and D.~R. {Stinebring}.
\newblock {Frequency-dependent Dispersion Measures and Implications for Pulsar
  Timing}.
\newblock {\em \apj}, 817:16, Jan 2016.

\bibitem{sro19}
Dan~R. {Stinebring}, Barney~J. {Rickett}, and Stella {Koch Ocker}.
\newblock {The Frequency Dependence of Scintillation Arc Thickness in Pulsar
  B1133+16}.
\newblock {\em \apj}, 870:82, Jan 2019.

\bibitem{gc18}
Gianfranco {Grillo} and James {Cordes}.
\newblock {Wave asymptotics and their application to astrophysical plasma
  lensing}.
\newblock {\em arXiv e-prints}, page arXiv:1810.09058, Oct 2018.

\end{thebibliography}

\end{document}